\documentclass[twocolumn,aps,superscriptaddress]{revtex4}
\usepackage{graphicx}
\usepackage{float}
\usepackage{dcolumn}
\usepackage{bm}
\usepackage{color}
\usepackage{amsmath}
\usepackage{amssymb}
\usepackage{amsfonts}
\usepackage{esint}
\usepackage{times}
\usepackage{xcolor}
\usepackage{phaistos}
\usepackage{braket}
\usepackage{comment}
\usepackage{multirow}

\usepackage{pdfpages}

\usepackage[colorlinks,linkcolor=blue,anchorcolor=blue,urlcolor=blue,urlcolor=blue,citecolor=blue]{hyperref}

\begin{document}

\title{Scalable native multiqubit gates via engineered noncomputational-state \\ interactions in superconducting fluxonium qubits}

\author{Peng Zhao}
\email{shangniguo@sina.com}
\affiliation{Quantum Science Center of Guangdong-Hong Kong-Macao Greater Bay Area, Shenzhen 518045, China}
\author{Peng Xu}
\affiliation{Institute of Quantum Information and Technology,
Nanjing University of Posts and Telecommunications, Nanjing, Jiangsu 210003, China}
\affiliation{State Key Laboratory of Quantum Optics Technologies and Devices, Shanxi University, Taiyuan, 030006, China}
\author{Zheng-Yuan Xue}
\email{zyxue83@163.com}
\affiliation{Key Laboratory of Atomic and Subatomic Structure and Quantum Control (Ministry of Education), Guangdong Basic Research Center of Excellence for Structure and Fundamental Interactions of Matter, and School of Physics, South China Normal University, Guangzhou 510006, China}
\affiliation{Guangdong Provincial Key Laboratory of Quantum Engineering and Quantum Materials, Guangdong-Hong Kong Joint Laboratory of Quantum Matter, and Frontier Research Institute for Physics, South China Normal University, Guangzhou 510006, China}
\affiliation{Quantum Science Center of Guangdong-Hong Kong-Macao Greater Bay Area, Shenzhen 518045, China}

\date{\today}

\begin{abstract}
Native multiqubit gates could be essential for bridging the gap from current noisy devices to future utility-scale quantum
computers, as they can substantially reduce circuit depth for near-term applications on noisy devices and may
also lower the physical overhead of fault-tolerant quantum computation. Here
we introduce a scalable protocol for implementing native multi-controlled gates on fluxonium qubits, supporting an arbitrary
number of control qubits ($N > 1$) while remaining compatible with existing single- and two-qubit gate
realizations. Our approach leverages engineered interactions in noncomputational state manifolds to enable qubit-state
selective transitions, which is activated for the direct implementation of $(C^{\otimes N})Z$ gates.
We show that in square lattices with fluxonium qubits, $CCZ$, $CCCZ$, and $CCCCZ$ gates with
errors around 0.01 (0.001) are achievable, with gate lengths of $50\,(100)\,\text{ns}$, $100\,(250)\,\text{ns}$,
and $150\,(300)\,\text{ns}$, respectively. Looking forward, integrating these native multi-controlled gates with primitive
single- and two-qubit gate sets within a single quantum processor could significantly enhance flexibility in circuit synthesis
and offer a promising alternative pathway toward utility-scale quantum computing.
\end{abstract}

\maketitle


\section{Introduction}\label{SecI}

Superconducting qubits~\cite{Kjaergaard2020,Gyenis2021} are a promising platform for quantum computing. Beyond the prevalent
transmon qubit~\cite{Koch2007}, fluxonium~\cite{Manucharyan2009} has emerged as a compelling alternative, demonstrating
long coherence times~\cite{Nguyen2019,Somoroff2023,Wang2025} and high-fidelity single- and two-qubit gate
operations~\cite{Ficheux2021,Bao2022,Xiong2022,Moskalenko2022,Dogan2023,Ding2023,Simakov2023,Ma2024,Zhang2024,Lin2025,Rower2024}, benefits
enabled by its strong anharmonicity and resilience to dielectric and flux
noise~\cite{Nguyen2019,Sun2023}. To advance toward utility-scale implementations, the key challenge is now to scale these
performance metrics through extensible coupling architectures and control methods~\cite{Moskalenko2022,Zhao2025,Barends2025}.

Generally, quantum computing can, in principle, be achieved using only single- and two-qubit gates, as
arbitrary gate operations can be decomposed into sequences of these elementary gates~\cite{Barenco1995}. This
principle holds across physical platforms, including superconducting qubits. However, many essential quantum algorithms are more
naturally expressed through native multi-qubit gates~\cite{Nielsen2000,Nam2020,Park2019}, whose decomposition
into these primitive gates often introduces significant overhead in circuit depth and gate
count~\cite{Nakanishi2024,Mandviwalla2018,Liang2021} and thus proves inefficient for practical implementation.
Consequently, native implementations of multi-qubit operations may offer substantial advantages for both
near-term applications~\cite{Bharti2022} and the long-term development of fault-tolerant quantum computing~\cite{Paetznick2013,Yoder2016,Chao2018,Tasler2025,Old2025}.

Despite these advantages and various proposed implementations~\cite{Zahedinejad2015,Zahedinejad2016,Barnes2017,Pedersen2019,Khazali2020,Feng2020,Zhao2020,Rasmussen2020,Rasmussen2020b,Gu2021,Baker2022,Glaser2023,Simakov2024,Zhao2025a}, the practical implementation and application of native multi-qubit gates in superconducting quantum computing
remain scarce~\cite{Song2017,Li2019,Roy2020,Kim2022,Warren2023,Hu2023,Itoko2024,Liu2025,Miao2025}. Generally,
this is due to two key challenges. First, while superconducting qubits intrinsically support multi-qubit interactions~\cite{Pedersen2019,Feng2020,Zhao2020,Glaser2023,Mezzacapo2014,Zhao2017,Sameti2017,Busnaina2025}, engineering even two-qubit interactions remains
difficult, e.g., even optimized two-qubit gates typically exhibit errors an order of magnitude higher than
single-qubit operations, see e.g., Ref.~\cite{Ding2023}. Extending this to three-qubit or larger gates is significantly
more challenging, as evidenced by the modest reported fidelities (typically $<99\%$ for three-qubit gates)~\cite{Song2017,Li2019,Roy2020,Kim2022,Warren2023,Hu2023,Itoko2024,Liu2025,Miao2025} and the scarcity of studies
on gates involving more than three qubits~\cite{Song2017}. These limitations mainly stem from control errors due to the
complexity of precise multi-qubit Hamiltonian engineering (e.g., engineering strong
three-body interactions while suppressing undesired two-body terms for three-qubit gates)~\cite{Li2019,Kim2022,Warren2023,Hu2023,Liu2025}
and incoherence from longer gate lengths (e.g., typically 250-350 ns for microwave-activated three-qubit gates)~\cite{Song2017,Kim2022,Warren2023,Itoko2024,Liu2025}. Second, the scalability of these multi-qubit gates to large-scale
systems and their compatibility with existing single- and two-qubit gate realizations remain uncertain. These constraints
currently restrict the practical utility of native multi-qubit operations and stifle the development of their
applications for quantum computing.

Here, we show that engineered interactions in noncomputational manifolds of superconducting fluxonium qubits can enable
scalable, native realization of $(C^{\otimes N})Z$ gates. This is achieved by engineering qubit-state-selective transitions
into noncomputational manifolds. The large frequency separation between the desired transition and other spurious transitions
allows selective activation of target multi-qubit interactions while suppressing unwanted terms. This approach provides
a path beyond existing schemes by enabling fast, scalable native multi-qubit gates without substantially increasing
calibration complexity, allowing for their co-integration with primitive single- and two-qubit gates in a unified
processor.

\section{circuit model and system Hamiltonian}\label{SecII}

\begin{figure}[tbp]
\begin{center}
\includegraphics[keepaspectratio=true,width=\columnwidth]{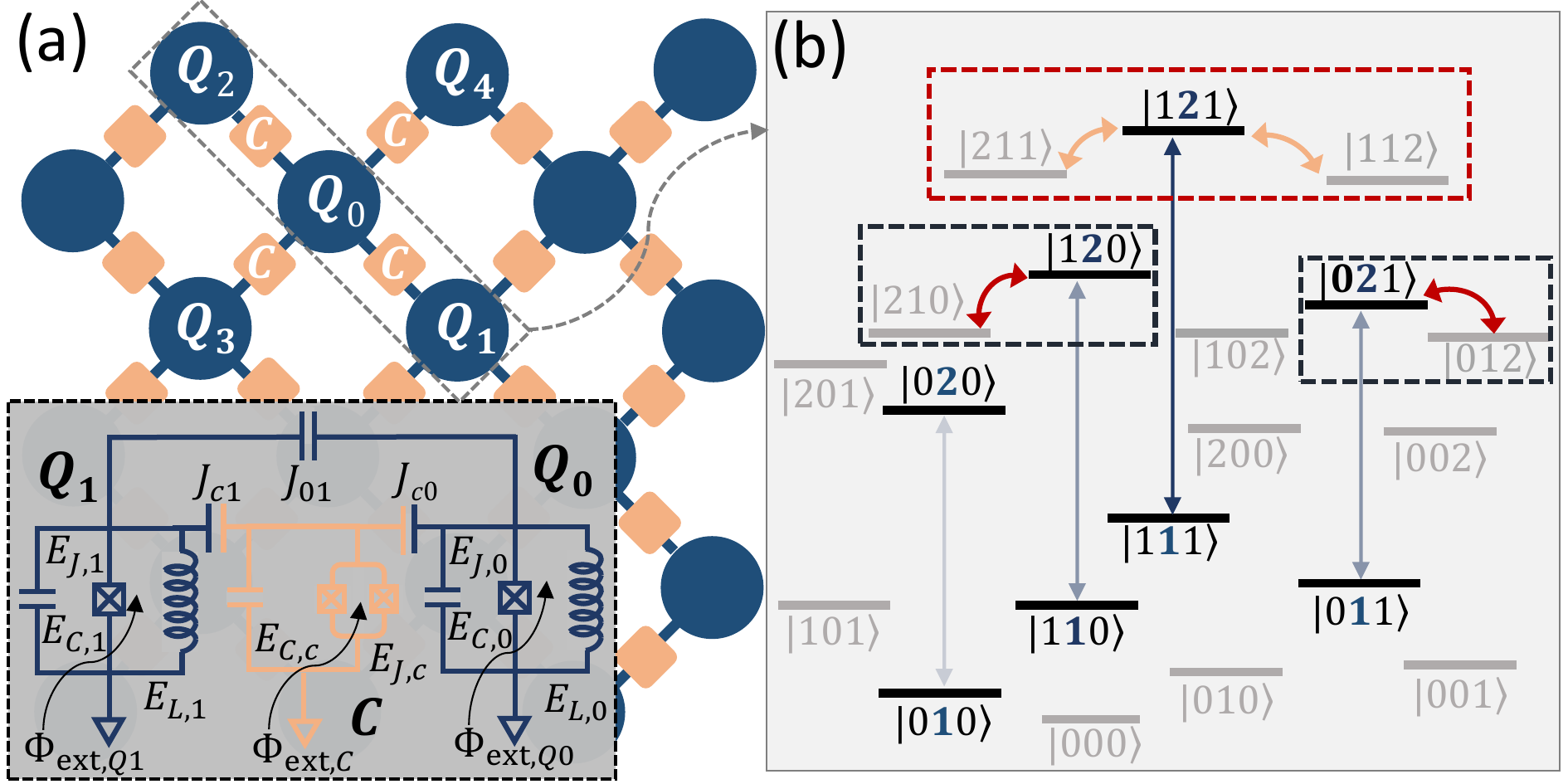}
\end{center}
\caption{(a) A 2D square qubit lattice comprising fluxoniums (circles) coupled via couplers (squares).
The inset depicts the fluxonium architecture featuring tunable plasmon interactions and qubit-state decoupling, where
fluxoniums are coupled via a transmon-based tunable coupler. (b) The energy levels of a
three-fluxonium system (with the system-state notation $|Q_{1},Q_{0},Q_{2}\rangle$), where $Q_{0}$ is coupled to two neighbors $Q_{1}$ and $Q_{2}$.
The coupler-mediated flip-flop interactions in the noncomputational-state manifolds (curved
arrows, $|210\rangle\leftrightarrow|120\rangle$, $|021\rangle\leftrightarrow|012\rangle$, and $|211(112)\rangle\leftrightarrow|121\rangle$) cause neighbor-state-dependent frequency shifts in $Q_{0}$'s plasmon transition $|1\rangle\leftrightarrow|2\rangle$ through interaction-induced
level repulsions.}
\label{fig1}
\end{figure}

\begin{table}[!htb]
\caption{\label{tab:circuit_parameters} Hamiltonian parameters of the fluxonium system.}
\begin{ruledtabular}
\begin{tabular}{cccc}
$ $&
$E_C$ (GHz)&
$E_L$ (GHz)&
$E_J$ (GHz)\\\hline
Fluxonium $Q_{0}$ & 1.41 & 0.80 & 6.27 \\
Fluxonium $Q_{1}$ & 1.30 & 0.59 & 5.71  \\
Fluxonium $Q_{2}$ & 1.33 & 0.60 & 5.40 \\
Fluxonium $Q_{3}$ & 1.35 & 0.63 & 5.60 \\
Fluxonium $Q_{4}$ & 1.35 & 0.70 & 5.60 \\
Transmon $C$ & 0.32 & $-$ & 55 \\
\hline
\hline
$ $ & $J_{c0 }$ (MHz) & $J_{c1}$ (MHz) & $J_{01}$ (MHz)\\\hline
Coupling strengths & 500 & 500  & 125
\end{tabular}
\end{ruledtabular}
\end{table}

\begin{table}[!htb]
\caption{\label{tab:qubit_parameters} Qubit parameters and coupler biases of the fluxonium system with
the circuit parameters listed in Table~\ref{tab:circuit_parameters}.}
\begin{ruledtabular}
\begin{tabular}{cccc}
$ $&
$\omega^{(01)}/2\pi$ (GHz)&
$\omega^{(12)}/2\pi$ (GHz)&
$\omega^{(03)}/2\pi$ (GHz)\\\hline
$Q_{0}$ & 0.298 & 5.621 & 8.347 \\
$Q_{1}$& 0.222  & 5.269 & 7.461  \\
$Q_{2}$ & 0.273 & 5.049 & 7.474 \\
$Q_{3}$ & 0.273 & 5.198 & 7.660 \\
$Q_{4}$ & 0.306 & 5.134 & 7.771 \\
$C$  & 11.537 & 11.194  & -\\
\hline
\hline
$(C^{\otimes N})Z$ & $N=1$ & $N=2$ & $N=3$ \\\hline
$\varphi_{\text{ext},cj}/2\pi$ & [0.413] & [0.413,0.420] & [0.403,0.420,0.410]
\end{tabular}
\end{ruledtabular}
\end{table}

For illustration purposes, the fluxonium qubit architecture that we study here is schematically
depicted in Fig.~\ref{fig1}(a), where fluxonium qubits are arranged in a 2D square lattice with
nearest-neighbor coupling mediated by transmon-based tunable couplers (inset). As
shown in Ref.~\cite{Zhao2025}, these couplers enable both qubit-state decoupling and tunable
plasmon interactions. For our theoretical framework (beyond the degree-4 connectivity
shown in Fig.~\ref{fig1}), we consider a general unit cell of large-scale qubit lattices comprising a central fluxonium $Q_{0}$
coupled to $N$ neighboring fluxoniums $(Q_{1},...,Q_{N})$, described by the Hamiltonian (hereafter, $\hbar=1$)
\begin{equation}
\begin{aligned}\label{eq1}
&H^{(N)}=\sum_{j=1}^{N}[H_{0,j}]+H_{0}
\\&H_{0,j}=H_{j}+[J_{c0}\hat n_0 \hat n_{cj}+J_{cj}\hat n_j \hat n_{cj}+J_{0j}\hat n_0 \hat n_{j}]
\\&\quad\quad \quad +[4 E_{C,cj} \hat n^2_{cj} - E_{J,cj}\cos(\frac{\varphi_{\text{ext},cj}}{2})\cos\hat\varphi_{cj}]
\end{aligned}
\end{equation}
with the fluxonium Hamiltonian $H_{k}=4 E_{C,k} \hat n^2_k + \frac{E_{L,k}}{2}(\hat\varphi_k - \varphi_{\text{ext},k})^2 - E_{J,k}\cos\hat\varphi_k$ ($k=0,j$)~\cite{Manucharyan2009,You2019}.
Here, $E_C$, $E_J$, and $E_L$ denote the charging, Josephson, and inductive energies, respectively,
subscripts $j$ and $cj$ label the fluxonium and the coupler, $J_{ck}$ and $J_{0j}$ represent the coupler-fluxonium and the direct
fluxonium-fluxonium coupling strength, and $\varphi_\text{ext}$ is the external phase bias with
$\varphi_\text{ext}=2\pi\Phi_\text{ext}/\Phi_0$. Hereafter, the lowest three
energy levels of each fluxonium are denoted by $\{|0\rangle,\,|1\rangle,\,|2\rangle\}$ and all fluxoniums
are biased at their half-flux quantum sweet spots ($\varphi_{\text{ext},k}=\pi$).

In the present architecture, the small transition dipole of the fluxonium qubit, together with its
relatively low qubit transition frequency, enables effective decoupling between the
computational subspaces and the couplers, leading to ZZ interactions suppressed
below $10\,\rm kHz$~\cite{Ding2023,Zhao2025}. In contrast, the transmon-like dipole of the fluxonium plasmon transition ($|1\rangle\leftrightarrow|2\rangle$) supports strong coupler-mediated interactions~\cite{Zhao2025}. 
Thus, here we focus on the plasmon interactions. Assuming that the coupling
system operates in the dispersive regime (where the plasmon-coupler detuning significantly exceeds
their coupling strength), we can derive an effective Hamiltonian by eliminating the direct fluxonium-coupler coupling terms
in Eq.~(\ref{eq1})~\cite{Zhao2025}. Projected onto the noncomputational manifold spanned by $\{|1\rangle,\,|2\rangle\}$, this yields the
approximate form (see Appendix~\ref{A} for details)
\begin{equation}
\begin{aligned}\label{eq2}
H_{\rm eff}^{(N)}=&\frac{\omega_{0}^{(12)}}{2}Z_{0}^{(12)}+\sum_{j=1}^{N}\left[\frac{\omega_{j}^{(12)}}{2}Z_{j}^{(12)}
+g_{0j}X_{0}^{(12)}X_{j}^{(12)}\right],
\end{aligned}
\end{equation}
where $\omega_{k}^{(12)}$ is the plasmon transition frequency of $Q_{k}$, $g_{0j}$ represents the strength of the
coupler-mediated flip-flop interaction between $Q_{0}$ and $Q_{j}$, and
$\{X^{(12)}=|1\rangle\langle 2|+|2\rangle\langle 1|,\,Z^{(12)}=|2\rangle\langle 2|-|1\rangle\langle 1|\}$ are the Pauli operators
defined in the $\{|1\rangle,\,|2\rangle\}$ subspace.

We further consider that the plasmon detuning between $Q_{0}$ and its neighbors $\Delta_{0j}=|\omega_{0}^{(12)}-\omega_{j}^{(12)}|$
is far larger than the coupling strength $g_{0j}$, i.e., operating in the dispersive regime, and the $N$ neighboring
fluxoniums are non-degenerate in plasmon frequency. This configuration suppresses plasmon excitation exchange among
fluxoniums while preserves dispersive interactions between $Q_{0}$ and its all neighbors, which
make $Q_{0}$'s plasmon frequency dependent on the collective state of its neighbors. This dependence becomes
explicit when diagonalizing the Hamiltonian in Eq.~(\ref{eq2}) and keeping only terms involving $Q_{0}$, yielding the
approximate form
\begin{equation}
\begin{aligned}\label{eq3}
H_{\rm eff,0}^{(N)}&=\frac{Z^{(12)}_{0}}{2}\left(\omega_{0}^{(12)}+\sum_{\overrightarrow{s}}[\delta_{\overrightarrow{s}}|\overrightarrow{s}\rangle\langle \overrightarrow{s}|]\right),
\end{aligned}
\end{equation}
where $\delta_{\overrightarrow{s}}\equiv\sum_{j}s_{j}\chi_{j}$ is the accumulated frequency shift contributing from
all $Q_{0}$'s neighbors, which are in state $|\overrightarrow{s}\rangle=|s_{1},...,s_j,...,s_{N}\rangle$ with $s_{j}=\{0,\,1\}$, and
$\chi_{j}\approx g_{0j}^{2}/\Delta_{0j}$ denotes the dispersive shift from the $Q_{0}$-$Q_{j}$ interaction.

Equation~(\ref{eq3}) demonstrates that the dispersive coupling to neighboring fluxoniums generates a state-dependent frequency shift in
$Q_{0}$'s plasmon transition. This shifts can be engineered for enabling state-selective activation
of transitions connecting computational and non-computational subspace and thus inducing non-trivial phase factors for
computational states. While such mechanism has been previously exploited for CZ gates in two-qubit
systems ($N=1$)~\cite{Chow2013,Nesterov2018}, here we generalize this to implement native $(C^{\otimes N})Z$
gates. Since transition selectivity (addressability) is ultimately limited by the shift
magnitude, which consequently constrains both gate speed and gate fidelity, we should first analyze the underlying
shift mechanism before proceeding to gate implementations, with particular focus on parameter optimization
for maximizing the shift and improving the selectivity.

Accordingly, we consider an example of a three-fluxonium system (with the state
notation $|Q_{1},Q_{0},Q_{2}\rangle$), see Fig.~\ref{fig1}(a), where
a central fluxonium $Q_{0}$ coupled to two neighbors $Q_{1}$ and $Q_{2}$ ($N=2$). Figure~\ref{fig1}(b) depicts the
system energy levels and demonstrates that the level repulsion from the interactions among noncomputational manifolds, i.e., $\{|210\rangle,\,|120\rangle\}$, $\{|021\rangle,\,|012\rangle\}$, and $\{|211\rangle,\,|121\rangle,\,|112\rangle\}$ (see
blue and red dashed boxes), causes the frequency shift in $Q_{0}$'s plasmon transition $|1\rangle\leftrightarrow|2\rangle$ (see
blue arrows). To be more specific, for two coupled energy levels (e.g., $|210\rangle\leftrightarrow|120\rangle$), the interaction-induced
repulsion results in an upward shift of the higher level (e.g., $|120\rangle$) and a downward shift of the lower
level (e.g., $|210\rangle$) with the magnitude depending on the coupling strength and the level detuning.

To ensure addressability in these state-dependent transitions, the central fluxonium $Q_{0}$'s plasmon
frequency can lie either above or below that of all neighbors, see Fig.~\ref{fig1}(b) (for the qubit lattice
shown in Fig.~\ref{fig1}(a), the frequency allocation of fluxonium plasmons in each row and column should follow a zigzag
pattern~\cite{Zhao2020}), preventing cancellation of shifts from competing level repulsions with different neighbors.
Moreover, while our above analysis assumes the dispersive limit, practical implementations generally need to enter into
non-dispersive regimes~\cite{Goerz2017,Sung2021,Zhao2021} to achieve sufficiently strong shifts for fast gates
while maintaining the validity of the approximate model given in Eq.~(\ref{eq3}) (while the model's validity, in theory, is
independent of the number of neighbors, it breaks down for practical large systems, see Appendix~\ref{B}).

\begin{figure}[tbp]
\begin{center}
\includegraphics[keepaspectratio=true,width=\columnwidth]{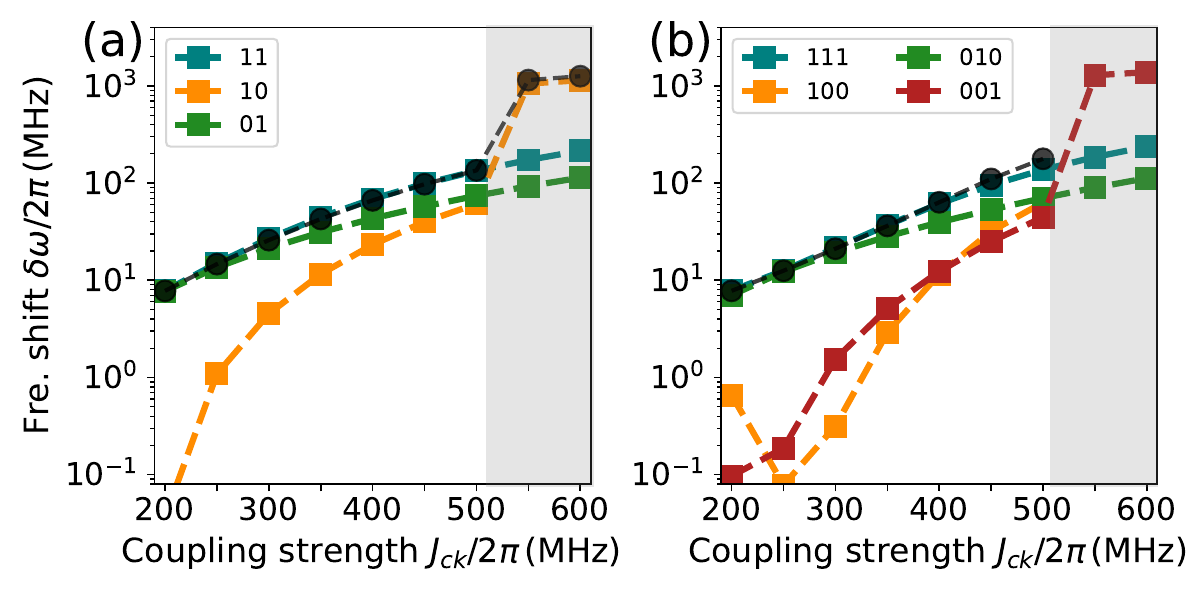}
\end{center}
\caption{The neighbor-state-dependent frequency shift in the central fluxonium versus the the coupler-fluxonium
coupling strength $J_{ck}$ for $N=2$ (a) and $N=3$ (b). Black circles denote the accumulated shift contributing from
all neighbors, i.e., $(\delta_{\protect\overrightarrow{01}}+\delta_{\protect\overrightarrow{10}})$ and $(\delta_{\protect\overrightarrow{001}}+\delta_{\protect\overrightarrow{010}}
+\delta_{\protect\overrightarrow{100}})$ are for $N=2$ and $N=3$, respectively. The gray regions show the
parameter space where the relation of $\delta_{\protect\overrightarrow{s}}\equiv\sum_{j}s_{j}\chi_{j}$ breaks
down (where the negative values of $\delta_{\protect\overrightarrow{100}}$ are excluded from the data presented
in (b)).}
\label{fig2}
\end{figure}

To validate the choice of operating regime and the approximate model, we consider the circuit parameters summarized in
Table~\ref{tab:circuit_parameters} with the corresponding qubit parameters listed in Table~\ref{tab:qubit_parameters}.
Figure~\ref{fig2}(a) shows the state-dependent
frequency shift versus the coupler-fluxonium coupling strength $J_{ck}$ for $N=2$ with the coupler biases (for tuning on
plasmon interactions) listed in Table~\ref{tab:qubit_parameters}. As $J_{ck}$ increases, the frequency shift grows
accordingly. The calculated shift $\delta_{\overrightarrow{11}}$ (teal squares) agrees well with the sum of individual neighbor
contributions $\delta_{\overrightarrow{01}}+\delta_{\overrightarrow{10}}$ (black circles) up to $J_{ck}/2\pi\approx 500\,\rm MHz$.
Beyond this point (gray shaded region), the relationship breaks down and shift magnitudes (e.g., $\delta_{\overrightarrow{10}}$
and $\delta_{\overrightarrow{100}}$) exhibit abrupt jumps. These behaviors are certainly expected for systems operating far outside the dispersive regime assumed in the derivation of Eqs.~(\ref{eq2}) and~(\ref{eq3}). Similar nonlinear effects also appear in the
four-qubit case ($N=3$), see Fig.~\ref{fig2}(b), and five-qubit systems ($N=4$), see Appendix~\ref{B}.

\section{Microwave-activated $(C^{\otimes N})Z$ gates}\label{SecIII}

\begin{figure*}[tbp]
\begin{center}
\includegraphics[width=15cm,height=6cm]{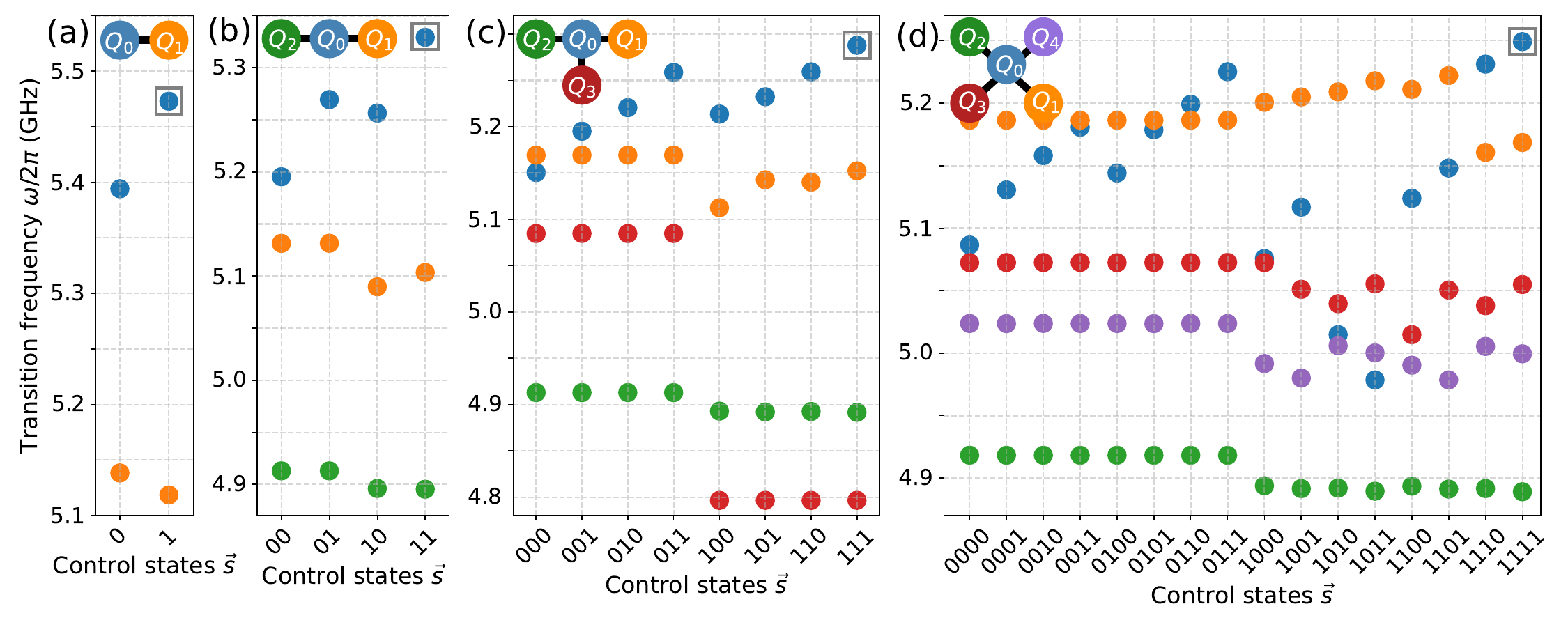}
\end{center}
\caption{State-dependent plasmon frequencies for each fluxonium under varying neighbor-state
configurations. The gray boxes highlight the gate frequency (i.e., the frequency of the central
fluxonium's $|1\rangle\leftrightarrow|2\rangle$ with all neighbors in state $|1\rangle$) for implementing
$(C^{\otimes N})Z$ gates. In (a-d), minimum detunings between the gate transition and other nearby transitions
are $(\delta^{'}_{\protect\overrightarrow{s}})_{\rm Min}/2\pi=\{-78.8,\,-59.7,\,-28.7,\,-18.0\}\, {\rm MHz}$ for $N=\{1,\,2,\,3,\,4\}$, respectively.}
\label{fig3}
\end{figure*}

\begin{figure}[tbp]
\begin{center}
\includegraphics[keepaspectratio=true,width=\columnwidth]{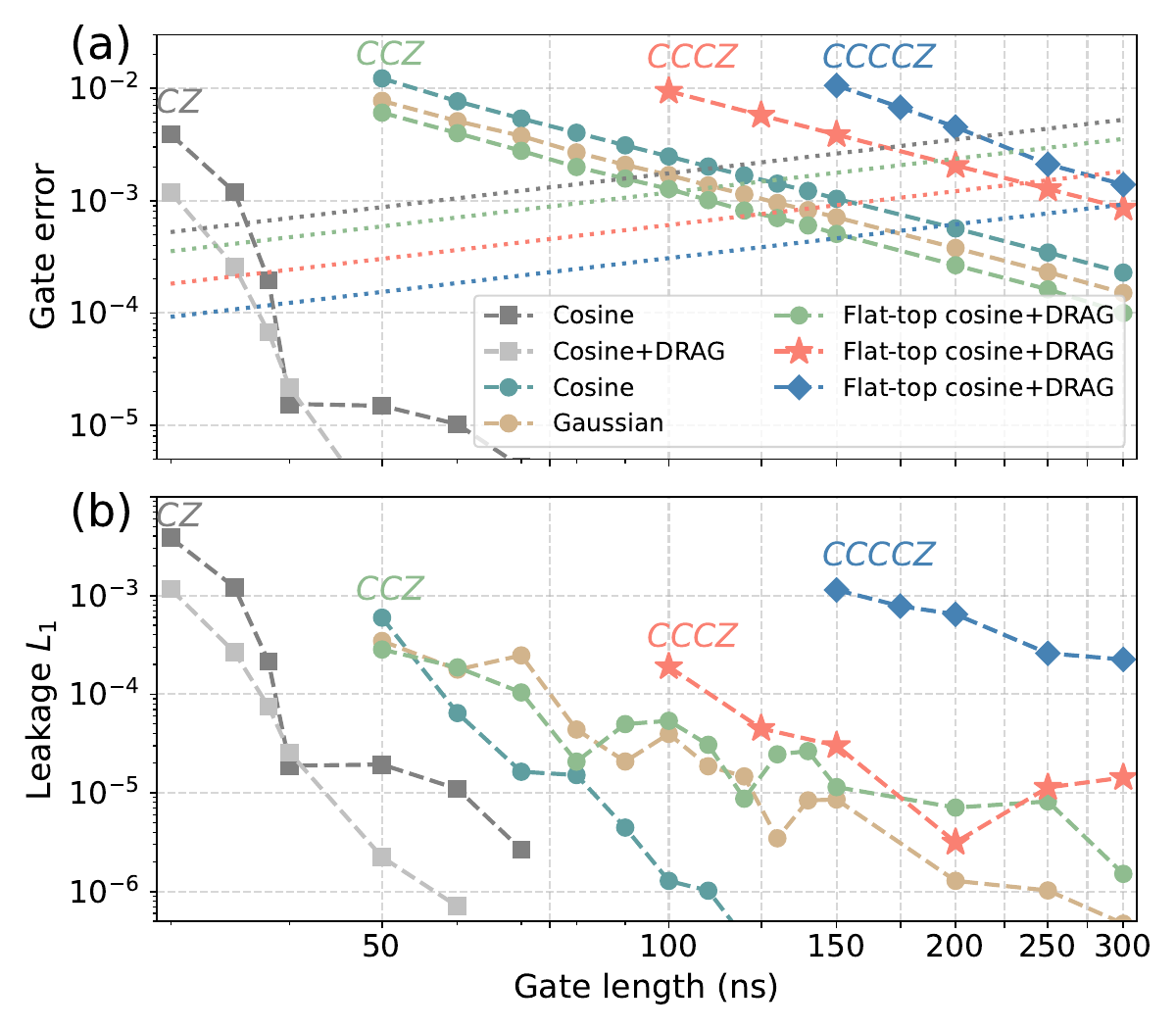}
\end{center}
\caption{Gate errors (a) and leakage (b) of $(C^{\otimes N})Z$ ($N=1,\,2,\,3,\,4$) versus gate lengths (excluding the coupler flux ramping) with different
drive pulse shapes. Here, the DRAG pulse parameters are chosen to suppress the spurious transition nearest to the target gate
transition, see Fig.~\ref{fig3}, and for $CCZ$, $CCCZ$, and $CCCCZ$ gates, the cosine ramp times of the flat-top pulses are $10$ ns, $20$ ns,
and $30$ ns, respectively. In (a), the calculated incoherence gate errors arising from  relaxation and dephasing of noncomputational gate
levels are represented by dotted lines in distinct colors: gray for $CZ$, green for $CCZ$, red for $CCCZ$, and blue for $CCCCZ$ gates.}
\label{fig4}
\end{figure}

Given the system parameters specified in Table~\ref{tab:circuit_parameters}, here we show that the shifts can
enable state-selective transitions in microwave-driven multi-qubit systems, thereby imprinting
non-trivial phases on computational states for $(C^{\otimes N})Z$ gate realizations. Figure~\ref{fig3} shows
the neighbor-state-dependent plasmon transition frequencies for all fluxoniums (including the central fluxonium) for
$N=\{1,\,2,\,3\}$. As demonstrated in pervious works~\cite{Ding2023,Chow2013,Nesterov2018} and also
shown in Fig.~\ref{fig3}(a) for $N=1$, large shifts can enable selective driving of one specific
transition (i.e., driving the target gate transition while maintaining suppression of all other transitions), e.g., the
$|11\rangle\leftrightarrow|21\rangle$ transition, thus inducing an extra phase accumulation on the $|11\rangle$ state
upon completion of a Rabi cycle. A controlled-Z ($CZ$) gate is thus realized (up to single-qubit Z rotations) when the
accumulated phase reaches $\pi$. Similarly, as shown in Figs.~\ref{fig3}(b) and~\ref{fig3}(c), the shift also enable
state-selective plasmon transitions in three- and four-fluxonium systems (as well as five-fluxonium
systems, see Fig.~\ref{fig3}(d) and Appendix~\ref{B}).

As operating in the non-dispersive regime for fast gates, the system exhibits strong state hybridization
within noncomputational manifolds. This generally enables the activation of any computational-to-noncomputational
transition with microwave drives applied to any fluxonium in the coupled system, regardless
of its site position. The gate speed is thus constrained by the minimal detuning between the target gate
transition and all other computational-to-noncomputational transitions, see Fig.~\ref{fig3}.
We therefore select the central fluxonium $Q_{0}$'s plasmon transition with the neighbor-state
configuration $|1\rangle^{\otimes N}$ (see grey boxes in Fig.~\ref{fig3}) as the target transition to
maximize gate speed and suppress spurious transitions.

Following Eq.~(\ref{eq3}), the dynamics with the microwave
drive (at the gate transition frequency) on the central fluxonium can be approximately described by the rotating-frame
Hamiltonian (with respect to the drive)
\begin{equation}
\begin{aligned}\label{eq4}
H_{\rm gate}^{(N)}=&\sum_{|\overrightarrow{s}\rangle\neq |1\rangle^{\otimes N}}
\left[(\frac{\delta^{'}_{\overrightarrow{s}}}{2}Z_{0}^{(12)}+\frac{\Omega}{2}X_{0}^{(12)})\otimes |\overrightarrow{s}\rangle\langle \overrightarrow{s}| \right]
\\& + \frac{\Omega}{2}X_{0}^{(12)}\otimes(|1\rangle\langle 1|)^{\otimes N},
\end{aligned}
\end{equation}
where $\Omega$ is the drive magnitude and $\delta^{'}_{\overrightarrow{s}}$ is the plasmon-drive detuning for $Q_{0}$ with
the neighbor-state configuration $|\overrightarrow{s}\rangle$. From Eq.~(\ref{eq4}), the drive activates both the target gate transition
$|1\rangle\otimes|1,...,1\rangle\leftrightarrow |2\rangle\otimes|1,...,1\rangle$ (the term in the second line) and
unwanted off-resonant transitions $|s_0\rangle\otimes|s_1,...,1,...,s_N\rangle\leftrightarrow |s_0\rangle\otimes|s_1,...,2,...,s_N\rangle$ (the
terms in the first line). By choosing the drive length $t_g$ to satisfy $\int_{0}^{t_g}\Omega dt=2\pi$ while neglecting
these off-resonant transitions, only the computational state $|1\rangle^{\otimes (N+1)}$ acquires a non-trivial
phase $\pi$, thus implementing the $(C^{\otimes N})Z =\text{diag}(1,1,1,...,1,-1)$ gate up to local Z rotations.

However, in practical implementations, off-resonant transitions can introduce non-negligible errors, particularly
when pursuing fast gates through strong drives in systems with many nearby
transitions. These spurious transitions primarily generate two error
types: leakage and unwanted phase accumulation. While leakage from dominant off-resonance transitions can be
mitigated through synchronization protocols~\cite{Ficheux2021} or pulse shaping
techniques (e.g., DRAG~\cite{Motzoi2009}), such methods show limited effectiveness in systems with numerous nearby
transitions. The phase, arising from ac-Stark shifts $\sim \Omega^{2}/4\delta^{'}_{\overrightarrow{s}}$ due to
off-resonant drives, comprise both single-qubit phases (correctable through local Z rotations) and problematic multi-qubit
conditional phases. For $(C^{\otimes N})Z$ gates, the off-resonant drives introduce $N+1$ single-qubit phases
and $C_{N+1}^{n}$ $n$-qubit conditional phases ($2\leq n\leq N$). Compensating for the latter typically requires
additional multi-qubit phase gates~\cite{Zhao2025a,Liu2025} or refocusing techniques~\cite{Glaser2023,Simakov2024}, undermining
the advantages of the native implementations compared to decomposed approaches and also complicating the gate
calibration. Thus, here the fundamental strategy for error suppression involves engineering plasmon
interactions to maximize the minimal state-dependent detuning $(\delta^{'}_{\overrightarrow{s}})_{\rm Min}$.

Accordingly, in our architecture, $(C^{\otimes N})Z$ gates can be realized by firstly turning on the plasmon interactions (i.e., tuning
couplers from their idle points $\varphi_{\text{ext},cj}/2\pi\approx 0$ to the interaction points $\varphi_{\text{ext},cj}/2\pi\approx 0.41$, see Table~\ref{tab:qubit_parameters}) for achieving large $(\delta^{'}_{\overrightarrow{s}})_{\rm Min}$, then applying a resonant
$2\pi$-pulse at the gate transition frequency, and finally returning couplers to their idle points. Here, we note that as in
Refs.~\cite{Nesterov2018,Ding2023,Zhao2025}, here we employ simultaneous microwave drives with identical amplitude and
frequency applied to both $Q_{0}$ and $Q_{1}$ for realizing target gates and thus reducing unwanted phase accumulations
from off-resonant transitions. Following tune-up procedures for CZ gates~\cite{Ding2023,Zhao2025}, drive
parameters (frequency and peak drive amplitude for a given pulse shape) for $CCZ$, $CCCZ$, and $CCCCZ$
gates are optimized by minimizing both leakage~\cite{Wood2018} and target conditional phase errors within the gate subspace, e.g., $\{|1\rangle\otimes|11\rangle,\,|1\rangle\otimes|01\rangle,\,|0\rangle\otimes|11\rangle,\,|0\rangle\otimes|01\rangle\}$ for $CCZ$
gates. We note that the choice of gate subspace remains flexible, with the key essential requirement being that it includes the states involved in the target gate transition. Therefore, for implementing a CCZ gate, an alternative valid gate subspace could be defined as, for example, $\{|1\rangle\otimes|11\rangle,\,|1\rangle\otimes|01\rangle,\,|1\rangle\otimes|10\rangle,\,|1\rangle\otimes|00\rangle\}$.

By considering different pulse shapes (see Appendix~\ref{C}), Figure~\ref{fig4}(a) shows $(C^{\otimes N})Z$ gate errors (up to local Z
rotations)~\cite{Pedersen2007} versus gate lengths (excluding the coupler flux ramping) for $N=\{1,\,2,\,3,\,4\}$ and
demonstrates that beyond high-fidelity $CZ$ gates, $CCZ$, $CCCZ$, and $CCCCZ$ gates with errors of approximately 0.01 (0.001) can also be achieved at gate
lengths of $50\,(100)$ ns, $100\,(250)$ ns, and $150\,(300)$ ns, respectively. Furthermore, assuming typical coherence times of $T_{1}^{21}=T_{\phi}^{21}=10\,\mu$s~\cite{Ficheux2021,Ding2023}, Figure~\ref{fig4}(a) also shows the incoherence gate errors resulting
from relaxation and dephasing of non-computational levels during gate operations, see Appendix~\ref{D}. These results indicate
that gate errors of $\sim 0.001$ could be achievable for $N=2,3,4$ under realistic experimental conditions.
When further examining the leakage~\cite{Wood2018} shown in Fig.~\ref{fig4}(b), a key difference emerges between CZ and
$(C^{\otimes N})Z$ ($N\geq2$), i.e., while CZ gate performance is predominantly limited by leakage from off-resonant
transitions, which can be effectively suppressed using DRAG pulses, $(C^{\otimes N})Z$ ($N\geq2$) gates are primarily constrained
by unwanted conditional phase accumulation, instead of the leakage. As mentioned before, this distinction arises fundamentally because: CZ gates only
involves a single two-qubit conditional phase, whereas $(C^{\otimes N})Z$ ($N\geq2$) gate implementations inherently accumulate unwanted multi-qubit
conditional phases (e.g., three, six, and ten two-qubit phases for $N=2,3,4$) through drive-induced
ac Stark shifts in addition to the desired multi-qubit phases~\cite{note2025}.

\section{conclusion}\label{SecV}

In conclusion, we  propose scalable native multi-qubit $(C^{\otimes N})Z$ gates on fluxonium qubits
by exploiting engineered interactions within non-computational state manifolds. We show that
microwave-activated, qubit-state-selective transitions facilitate targeted multi-qubit
interactions while intrinsically suppressing spurious terms. This mechanism enables
fast, high-fidelity gates without a significant increase in calibration complexity.
This scalable approach, which combines high performance with simplified tune-up procedures
and remains fully compatible with existing single- and two-qubit gates, allows for the
integration of native multi-controlled gates with primitive gate sets within a
single quantum processor. We anticipate that quantum processors supporting both high-fidelity
native multi-qubit gates and primitive gates could benefit both near-term applications, such as
algorithm co-design favoring native multi-qubit interactions, and quantum error correction
by improving circuit efficiency and reducing overhead.

\begin{acknowledgments}
Peng Zhao would like to thank Zhuang Ma, Teng Ma, Meng-Jun Hu, Kunzhe Dai, and Ruixia Wang for insightful
discussions, Yanqing Zhu, Xiaotong Ni, Jia Liu, and Ling-Zhi Tang for their generous support. This work is supported by the
National Natural Science Foundation of China (Grants No.12204050 and No.92576110) and the Guangdong Provincial Quantum
Science Strategic Initiative (Grant No. GDZX2203001). Peng Xu is supported by the National Natural Science
Foundation of China (Grants No.12105146 and No.12175104) and the Program of State Key Laboratory of Quantum Optics
Technologies and Devices (No.KF202505).
\end{acknowledgments}

\appendix


\section{Effective system Hamiltonian}\label{A}

Following the approach outlined in Ref.~\cite{Zhao2025}, here we derive the effective system Hamiltonian presented in
the main text. Without loss of generality, we consider the two-qubit unit cell consisting of a central fluxonium qubit
$Q_{0}$ coupled to a neighboring fluxonium $Q_{1}$, with their interaction mediated by a transmon-based coupler. The
full system Hamiltonian takes the form:
\begin{equation}
\begin{aligned}\label{eqA1}
H^{(1)}=& \sum_{k=0,1}[4 E_{C,k} \hat n^2_k + \frac{E_{L,k}}{2}(\hat\varphi_k - \varphi_{\text{ext},k})^2 - E_{J,k}\cos\hat\varphi_k]\\
&+J_{c0}\hat n_0 \hat n_{c1}+J_{c1}\hat n_1 \hat n_{c1}+J_{01}\hat n_0 \hat n_{1}
\\&+4 E_{C,c1} \hat n^2_{c1} - E_{J,c1}\cos(\frac{\varphi_{\text{ext},c1}}{2})\cos\hat\varphi_{c1}.
\end{aligned}
\end{equation}
By approximating the transmon coupler as a weakly anharmonic oscillator~\cite{Koch2007} and introducing
\begin{equation}
\begin{aligned}\label{eqA2}
&\hat \varphi_{c1} = \phi_{c1,{\rm zpf}}(\hat a_{c1}^{\dag}+\hat a_{c1}),
\quad \hat n_{c1} = i n_{c1,{\rm zpf}}(\hat a_{c1}^{\dag}-\hat a_{c1})
\end{aligned}
\end{equation}
with
\begin{equation}
\begin{aligned}\label{eqA3}
&\varphi_{c1,{\rm zpf}} =\frac{1}{\sqrt{2}}\left[\frac{8E_{C,c1}}{E_{J,c1}(\varphi_{\text{ext},c1})}\right]^{\frac{1}{4}},
\\&n_{c1,{\rm zpf}}= \frac{1}{\sqrt{2}}\left[\frac{E_{J,c1}(\varphi_{\text{ext},c1})}{8E_{C,c1}}\right]^{\frac{1}{4}},
\end{aligned}
\end{equation}
where $a_{c1}$ ($a_{c1}^{\dag}$) denotes the destroy (creation) operator and $\phi_{c1,{\rm zpf}}$ ($n_{c1,{\rm zpf}}$) represents
the phase (number) zero-point fluctuation, the coupler Hamiltonian can be approximated by
\begin{equation}
\begin{aligned}\label{eqA4}
H_{coupler}=\omega_{c1}\hat a_{c1}^{\dag}\hat a_{c1}+\frac{\alpha_{c1}}{2}\hat a_{c1}^{\dag}\hat a_{c1}^{\dag}\hat a_{c1}\hat a_{c1}.
\end{aligned}
\end{equation}
Here, the coupler transition frequency and the coupler anharmonicity are $\omega_{c1}=\omega_{p1}(1-3\lambda_{1}-9\lambda_{1}^{2})$
and $\alpha_{c1}=-\omega_{p1}(3\lambda_{1}+\frac{162}{8}\lambda_{1}^{2})$~\cite{Khezri2018}, respectively, with
\begin{equation}
\begin{aligned}\label{eqA5}
\omega_{p1}=\sqrt{8E_{J,c1}(\varphi_{\text{ext},c1})E_{C,c1}},\,\lambda_{1}=\frac{1}{3}\sqrt{\frac{E_{C,c1}}{8E_{J,c1}(\varphi_{\text{ext},c1})}}.
\end{aligned}
\end{equation}

Focusing on the fluxonium's plasmon transition of $|1\rangle\leftrightarrow|2\rangle$, the system Hamiltonian
becomes (redefining $a=-ia$ and $a^{\dag}=ia^{\dag}$)
\begin{equation}
\begin{aligned}\label{eqA6}
H_{12}^{(1)}=&\sum_{k=0,1}\left[\frac{\omega_{k}^{(12)}}{2}Z_{k}^{(12)}+g_{12,k}X_{k}^{(12)}(\hat a_{c1}+\hat a_{c1}^{\dag})\right]
\\&+H_{coupler}+g_{12,01}X_{0}^{(12)}X_{1}^{(12)},
\end{aligned}
\end{equation}
where $\{X^{(12)}=|1\rangle\langle 2|+|2\rangle\langle 1|,\,Z^{(12)}=|2\rangle\langle 2|-|1\rangle\langle 1|\}$ are the Pauli operators
defined in the $\{|1\rangle,\,|2\rangle\}$ subspace, $g_{12,k}\equiv J_{ck} n_{c1,{\rm zpf}}\langle 1| n_k |2\rangle$
and $g_{12,01}=J_{01}\langle 1| n_0 |2\rangle \langle 1| n_1 |2\rangle$ denote the coupling strength of the plasmon-coupler
interaction and the direct plasmon-plasmon interaction, respectively.

Considering that the system operates in the dispersive regime, i.e., the interaction strength $g_{12,k}$ significantly smaller
than the fluxonium-coupler detuning $\Delta_{12,k}=|\omega_{k}^{(12)}-\omega_{c1}|$, an effective Hamiltonian (to second
order in $g_{12,k}/\Delta_{12,k}$) can be obtained by eliminating the direct plasmon-coupler
interactions~\cite{Zueco2009,Yan2018}, leading to (recovering the formula shown in Eq.~(\ref{eq2}) of the main text with $N=1$)
\begin{equation}
\begin{aligned}\label{eqA7}
&H_{{\rm eff}}^{(1)}= \sum_{k=0,1}\left[\frac{\omega_{k}^{(12)}}{2}Z_{k}^{(12)}\right]+g_{01}X_{0}^{(12)}X_{1}^{(12)},
\\&g_{01}= g_{12,01}+\frac{g_{12,0}g_{12,1}}{2}\left[ \sum_{k=0,1}(\frac{1}{\Delta_{12,k}}-\frac{1}{S_{12,k}})\right],
\end{aligned}
\end{equation}
with $S_{12,k}=\omega_{k}^{(12)}+\omega_{c1}$. For clarity, here we omit the renormalization of the plasmon transition
frequency due to the plasmon-coupler interactions.

Further assuming that the plasmon-plasmon detuning $\Delta_{01}=|\omega_{0}^{(12)}-\omega_{1}^{(12)}|$ is far larger than
the plasmon-plasmon coupling strength $g_{01}$, i.e., $Q_{0}$ and $Q_{1}$ are coupled dispersively, Equation~(\ref{eqA7})
can be further simplified by eliminating the plasmon-plasmon coupling term (to second order
in $g_{01}/\Delta_{01}$), i.e.,
\begin{equation}
\begin{aligned}\label{eqA8}
&H_{{\rm eff}}^{(1)}=\frac{Z_{0}^{(12)}}{2}(\omega_{0}^{(12)}+\chi_{1})+\frac{Z_{1}^{(12)}}{2}(\omega_{1}^{(12)}-\chi_{1}),
\end{aligned}
\end{equation}
with $\chi_{1}=g_{01}^{2}/\Delta_{01}$ denoting the dispersive shift. Note that the above equation is implicitly
defined in the two-fluxonium subspace spanned by $\{|11\rangle,|12\rangle,|21\rangle,|22\rangle\}$. This means that
the dispersive shift in one fluxonium's plasmon transition becomes active only when its coupled neighbor is in
excited states. Consequently, when restricting to $Q_{0}$ and projecting onto $Q_{1}$'s computational basis
${|0\rangle,|1\rangle}$, the dispersive Hamiltonian reduces to:
\begin{equation}
\begin{aligned}\label{eqA9}
&H_{{\rm eff},0}^{(1)}=\frac{Z_{0}^{(12)}}{2}(\omega_{0}^{(12)}+\chi_{1}|1\rangle\langle 1|_{1}).
\end{aligned}
\end{equation}
This recovers the formula shown in Eq.~(\ref{eq3}) of the main text with $N=1$.

\begin{figure}[htbp]
\begin{center}
\includegraphics[keepaspectratio=true,width=\columnwidth]{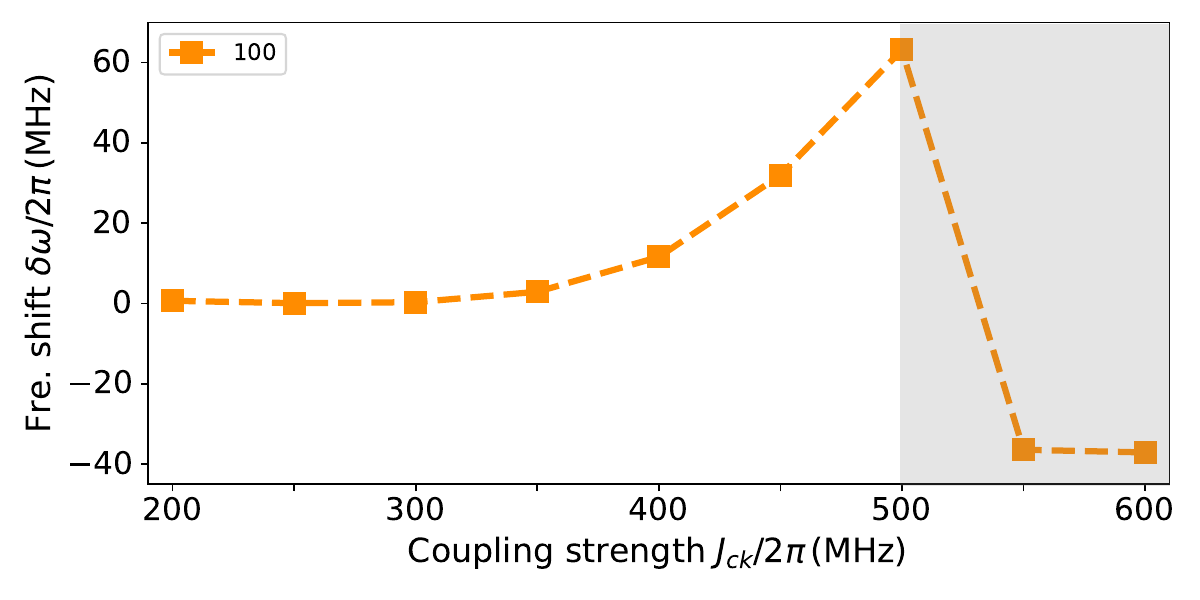}
\end{center}
\caption{The neighbor-state-dependent frequency shift $\delta_{\protect\overrightarrow{100}}$ versus the the coupler-fluxonium
coupling strength $J_{ck}$ for $N=3$. The gray regions show the parameter space where the approximate model breaks
down, characterized by $\delta_{\protect\overrightarrow{100}}$ undergoing discontinuous transitions between positive
and negative values. Here the coupler biases are listed in Table~\ref{tab:qubit_parameters} (namely [0.403, 0.420, 0.410]).}
\label{figS1}
\end{figure}

\begin{figure}[htbp]
\begin{center}
\includegraphics[keepaspectratio=true,width=\columnwidth]{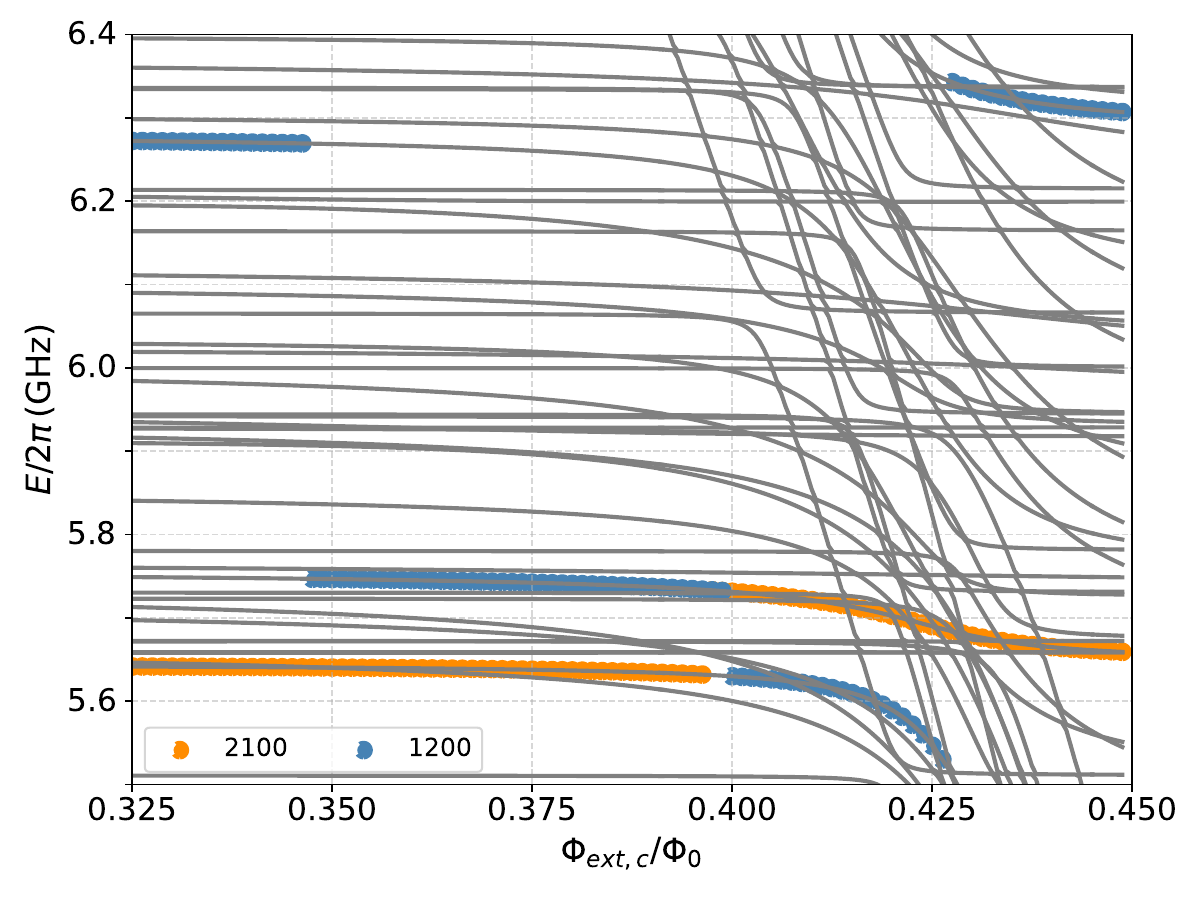}
\end{center}
\caption{Energy levels of the four coupled fluxonium system ($N=3$) versus the coupler flux bias $\varphi_{\text{ext},c3}$. 
Here, the the coupler-fluxonium coupling strength is $J_{ck}/2\pi=500\,\rm MHz$ and the other two couplers are biased at $\varphi_{\text{ext},c1(c2)}/2\pi=0.403\,(0.42)$.}
\label{figS2}
\end{figure}

\section{Extendability of $(C^{\otimes N})Z$ gate implementations}\label{B}

In theory, the validity of the approximate model is independent of neighbor count, especially for
systems operated in the dispersive regime. However, achieving large state-dependent shifts and fast
gates typically requires operating in strong non-dispersive regimes, necessitating careful parameter
optimization to balance shift magnitude with model validity. For the circuit parameters in
Table~\ref{tab:circuit_parameters}, Figure~\ref{fig3} (see the main text) reveals model breakdown (evidenced by the shift
magnitude undergoing abrupt jumps) when $J_{ck}/2\pi\geq 500\,\rm MHz$ for $N=2,3$. This is further
demonstrated in the four-qubit system ($N=3$) by the discontinuous sign transitions of $\delta_{\protect\overrightarrow{100}}$
with increasing $J_{ck}$, see Fig.~\ref{figS1} with the coupler biases listed in
Table~\ref{tab:qubit_parameters} (namely [0.403, 0.420, 0.410]). By examining the system spectrum, we find that these discontinuous
sign transitions result from on-resonant plasmon coupling between $Q_{0}$ and $Q_{1}$. More specifically, Fig.~\ref{figS2} shows
the system spectrum as a function of one coupler bias while the other two are fixed at [0.403, 0.420]. Near a coupler bias
of about 0.400, a clear avoided crossing appears between the states $|2100\rangle$ and $|1200\rangle$, corresponding precisely
to the plasmon coupling between $Q_{0}$ and $Q_{1}$.

\begin{figure}[htbp]
\begin{center}
\includegraphics[keepaspectratio=true,width=\columnwidth]{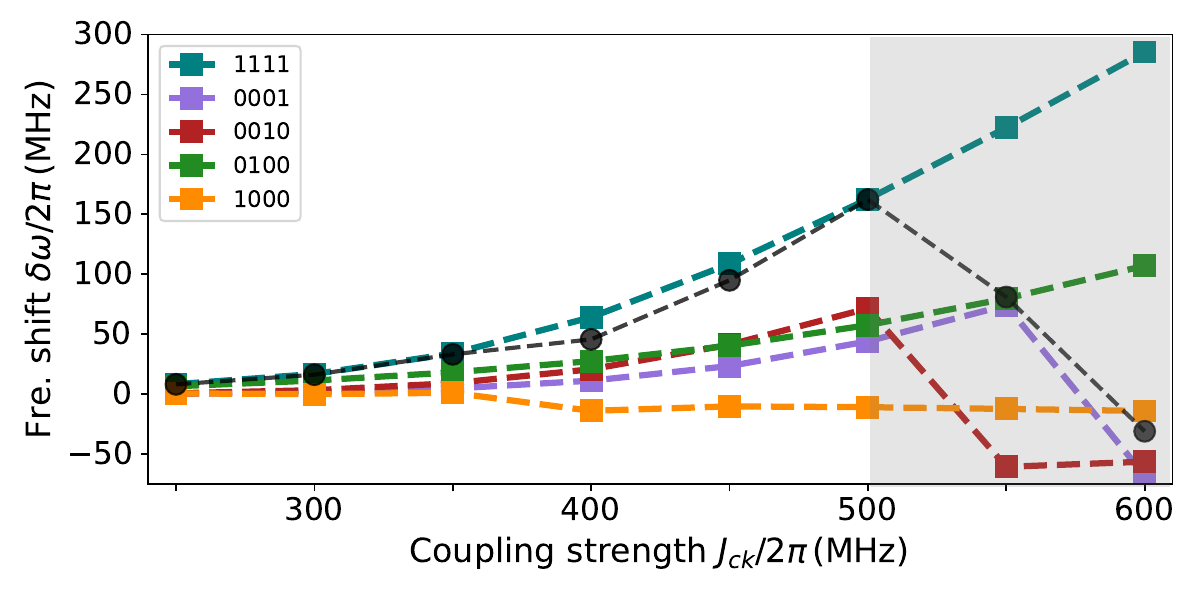}
\end{center}
\caption{The neighbor-state-dependent frequency shift in the central fluxonium versus the the coupler-fluxonium
coupling strength $J_{ck}$ for $N=4$. Black circles denote the accumulated shift contributing from
all neighbors, i.e., $(\delta_{\protect\overrightarrow{0001}}+\delta_{\protect\overrightarrow{0010}}
+\delta_{\protect\overrightarrow{0100}}+\delta_{\protect\overrightarrow{1000}})$. The gray regions indicate the
parameter space where the relation of $\delta_{\protect\overrightarrow{s}}\equiv\sum_{j}s_{j}\chi_{j}$
breaks down.}
\label{figS3}
\end{figure}

\begin{figure}[tbp]
\begin{center}
\includegraphics[keepaspectratio=true,width=\columnwidth]{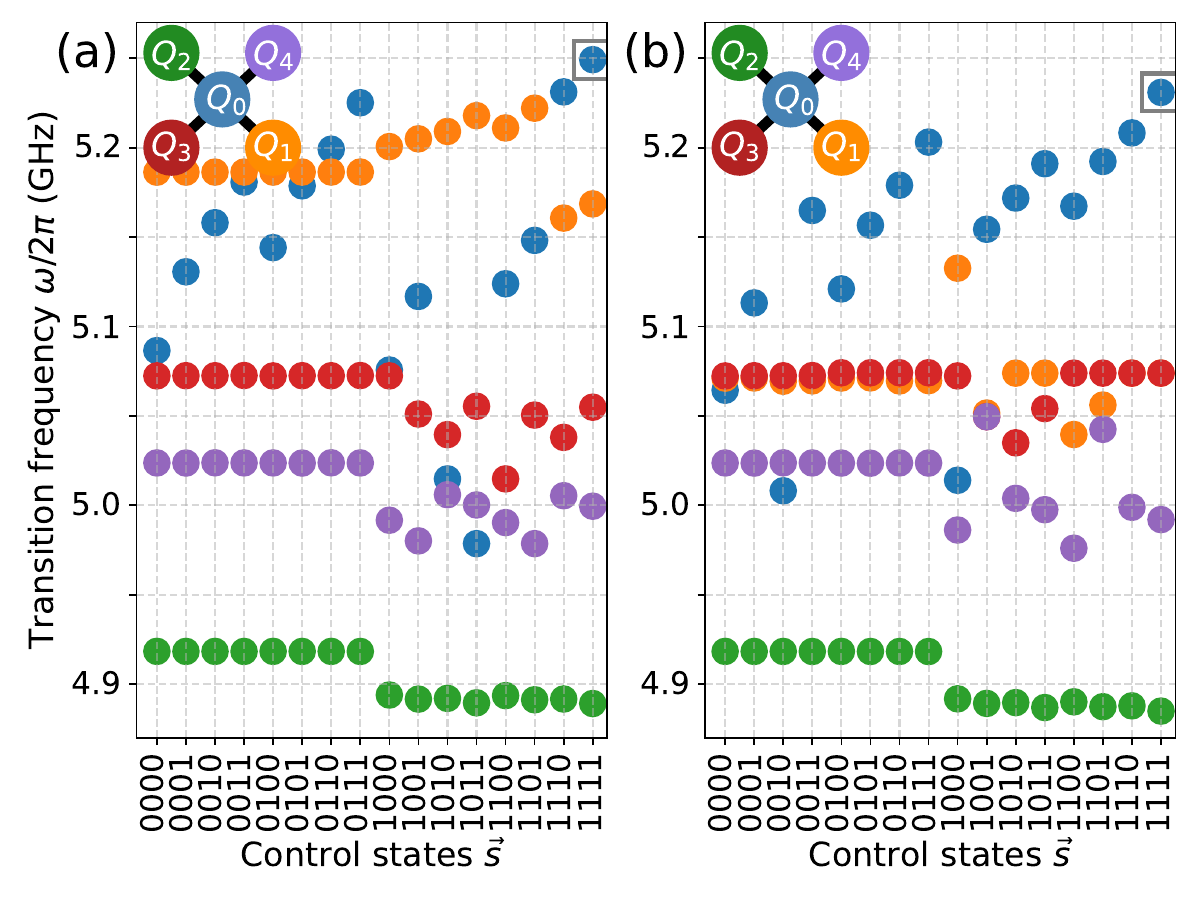}
\end{center}
\caption{State-dependent plasmon transition frequencies for each fluxonium under varying neighbor-state
configurations with $N=4$. (a) The gray box highlights the gate frequency for implementing
$CCCCZ$ gates with the minimum state-dependent detuning of $(\delta^{'}_{\protect\overrightarrow{s}})_{\rm Min}/2\pi=-18.0\,\rm MHz$
and the coupler biases at their interaction configurations are $\varphi_{\text{ext},cj}/2\pi=[0.395,0.419,0.413,0.411]$.
(b) The circuit parameters are same as in (a), except for $Q_{1}$'s $E_{J}$, which is adjusted to $5.60\,\rm GHz$.
This modification yields the modified fluxonium transition
frequencies $\{\omega_{1}^{(01)},\omega_{1}^{(12)},\omega_{1}^{(03)}\}/2\pi=\{0.223,5.83,7.422\}\,\rm GHz$ with increased
plasmon detuning between $Q_{0}$ and $Q_{1}$ . Here, the coupler biases are
$\varphi_{\text{ext},cj}/2\pi=[0.410,0.419,0.413,0.411]$, leading to $(\delta^{'}_{\protect\overrightarrow{s}})_{\rm Min}/2\pi=-22.7\,\rm MHz$.}
\label{figS4}
\end{figure}

Analogous discontinuities in frequency shifts also appear in five-qubit systems ($N=4$, see Fig.~\ref{figS3}), where similar
level-repulsion-induced collisions occur, as shown in Fig.~\ref{figS4}(a). Figure~\ref{figS4}(b) shows that such
collisions can be partially mitigated by increasing the $Q_{0}$-$Q_{1}$ plasmon detuning. We thus expect that level-repulsion
induced collisions can be more frequent in larger systems with crowding spectrums.

Moreover, although the approximate model predicts that the magnitude of the minimum state-dependent detuning
$(\delta^{'}_{\protect\overrightarrow{s}}){\rm Min}$ depends solely on the plasmon detuning $\Delta_{k}$ and coupling
strengths $g_{0k}$ (independent of neighbor count), one can expect that practical coupling systems operating in non-dispersive
regimes with strong state hybridization should exhibit a clear neighbor-dependent reduction
in $(\delta^{'}_{\protect\overrightarrow{s}}){\rm Min}$. As demonstrated in Figs.~\ref{fig3} and~\ref{figS4}(a), the detuning
magnitude decreases progressively from $-78.8$ MHz ($N=1$) to $-18.0$ MHz ($N=4$) with increasing
neighbor number. Furthermore, the substantial capacitive coupling required ($\sim 500$ MHz) between fluxoniums and
transmon-based couplers introduces significant capacitive loading~\cite{Ding2023,Rosenfeld2024}, given the typical $\sim 1$ GHz charge energies
of fluxonium qubits. This physical constraint makes high-connectivity architectures ($N>4$) based on fluxoniums particularly
challenging to implement~\cite{Rosenfeld2024}. Consequently, all these considerations suggest that realizing fast $(C^{\otimes N})Z$ gates
for $N>4$ presents non-trivial engineering challenges that require careful system design.

\begin{figure*}[tbp]
\begin{center}
\includegraphics[width=18cm,height=10cm]{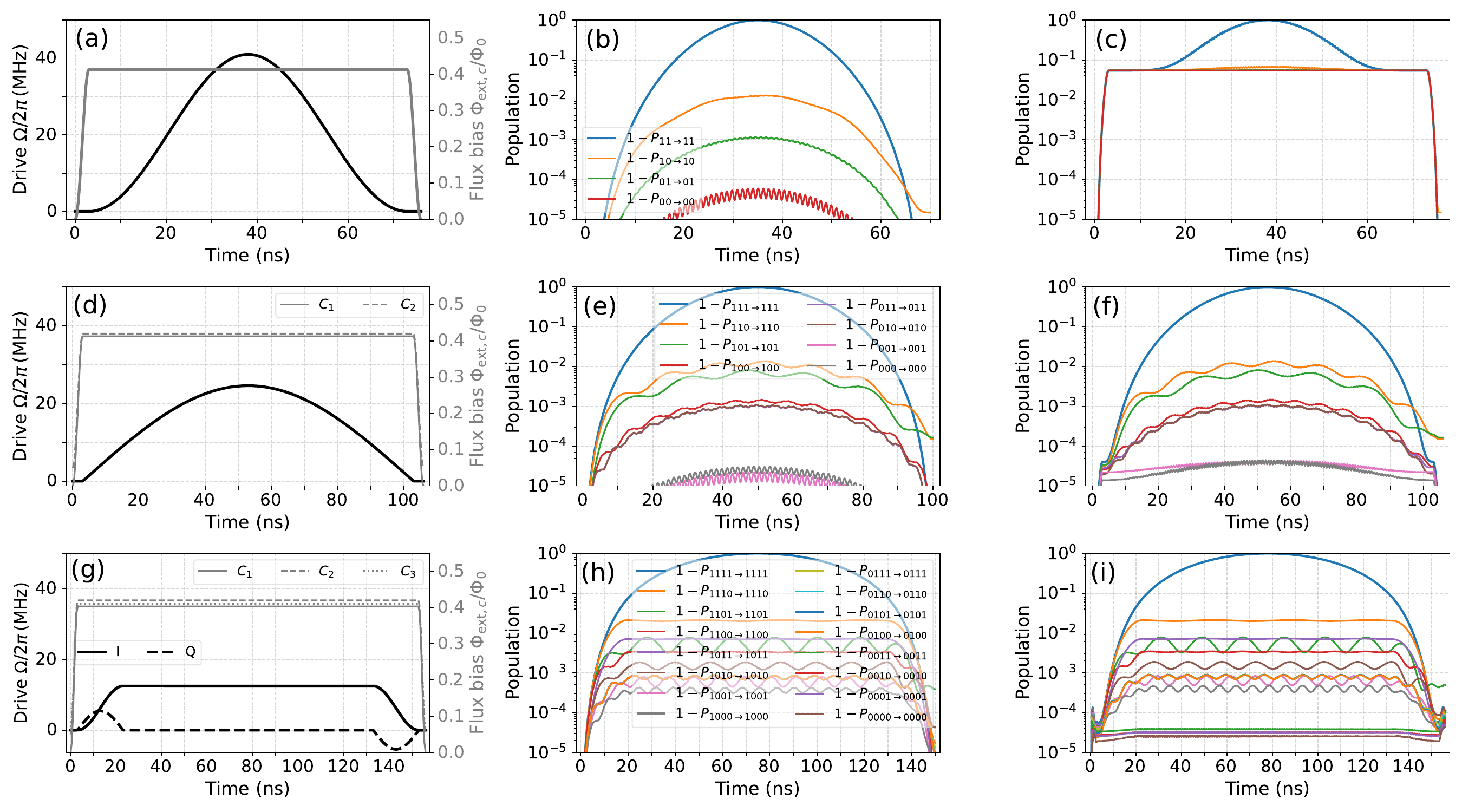}
\end{center}
\caption{(a,d,g) Typical control pulse shapes (Cosine, Gaussian, and Flat-top cosine with DRAG correction)
alongside (b,c; e,f; h,i) the corresponding gate dynamics for microwave-activated $(C^{\otimes N})Z$ gates, comparing
implementations without (b,e,h) and with (c,f,i) coupler flux ramping. In (a,d,g), grey lines denote the
flux pulses (flat-top cosine) for biasing the coupler during gates with the ramp time of 3 ns. (a,b,c), (d,e,f), and (g,h,i) are
for $N=1,2,3$, respectively. $P_{s\rightarrow s}$ denotes the population in $|s\rangle$ with the system
initialized in the state of $|s\rangle$.}
\label{figS5}
\end{figure*}

\section{Microwave-activated $(C^{\otimes N})Z$ gates}\label{C}

Following approaches for $CZ$ gate realization~\cite{Nesterov2018,Ding2023,Zhao2025}, here we employ simultaneous microwave drives
with identical amplitude and frequency applied to both $Q_{0}$ and $Q_{1}$ for realizing target gates and
reducing unwanted phase accumulations from off-resonant transitions. The corresponding driven Hamiltonian takes the form:
\begin{equation}
\begin{aligned}\label{eqC1}
H_{d}=\sum_{k=0,1}A(t)\cos(\omega_{d}t+\phi_{k})\hat n_{k},
\end{aligned}
\end{equation}
where $A$, $\omega_{d}$, and $\phi_{k}$ denotes the amplitude, frequency, and the
phase of the drive, respectively. Note that the relative phase between the two drives
are determined by maximizing constructive interference at the target gate transition~\cite{Ding2023}.

As shown in Fig.~\ref{figS5}, we consider using three types of pulses for driving gate
transitions, i.e., (i) \emph{Cosine}, which is defined by
\begin{equation}
\begin{aligned}\label{eqC2}
&A(t)=\Omega_{d}\left(1-\cos\frac{2\pi t}{t_{g}}\right),
\end{aligned}
\end{equation}
where $\Omega_{d}$ denotes the peak drive amplitude and $t_g$ represents the gate length (excluding
the coupler bias ramping); (2) \emph{Gaussian}, which has the following form
\begin{equation}
\begin{aligned}\label{eqC3}
&A(t)=\Omega_{d}\frac{e^{\frac{-(t-t_g/2)^2}{2\sigma^2}}-e^{\frac{-(t_{g}/2)^2}{2\sigma^2}}}{1-e^{\frac{-(t_{g}/2)^2}{2\sigma^2}}}.
\end{aligned}
\end{equation}
with $\sigma=t_g/2$; (iii) \emph{Flat-top cosine}, that is given as
\begin{align}
A(t)\equiv
\begin{cases}
\Omega_{d}\frac{1-\cos{(\pi \frac{t}{t_r}})}{2}  \;, &0<t<t_r\\
\Omega_{d}\;,  &t_r<t<t_g-t_r\\
\Omega_{d}\frac{1-\cos{(\pi \frac{t_g-t}{t_r}})}{2} \;, &t_g-t_r<t<t_g
\end{cases}
\label{eqC4}
\end{align}
with the ramp time $t_{r}$, and the full DRAG pulse is $A_{DRAG}(t)=A(t)+i\frac{\alpha}{(\delta^{'}_{\overrightarrow{s}})_{\rm Min}} \dot{A}(t)$ with $\alpha=1$
for minimizing leading leakage sources. Figures.~\ref{figS5}(a,d,g) also shows the typical flux pulse for basing the couplers, i.e., tuning
the coupler from their idle configurations to the interaction points.

As mentioned in the main text, the gate parameters (i.e., the drive peak amplitude $\Omega_{d}$ and the
drive frequency $\omega_{d}$) are optimized by minimizing both leakage~\cite{Wood2018} and target conditional
phase errors within the gate subspace given as $\{|1\rangle\otimes|11\rangle,\,|1\rangle\otimes|01\rangle,\,|0\rangle\otimes|11\rangle,\,|0\rangle\otimes|01\rangle\}$ for $N=2$,
$\{|1\rangle\otimes|111\rangle,\,|1\rangle\otimes|011\rangle,\,|0\rangle\otimes|111\rangle,\,|0\rangle\otimes|011\rangle\}$ for $N=3$, and
$\{|1\rangle\otimes|1111\rangle,\,|1\rangle\otimes|0111\rangle,\,|0\rangle\otimes|1111\rangle,\,|0\rangle\otimes|0111\rangle\}$ for $N=4$.
Given the optimized gate parameters, Figure~\ref{figS5} shows the typical gate dynamics.
Notably, comparisons between implementations without (b,e,h) and with (c,f,i) coupler flux ramping reveal negligible population
differences (at the level of $\sim 10^{-4}$). This is consistent with expectations given the near-complete
decoupling of the computational subspace and the coupler that effectively suppresses non-adiabatic transitions for
computational states during coupler flux ramping~\cite{Zhao2025}.

\begin{figure}[tbp]
\begin{center}
\includegraphics[keepaspectratio=true,width=\columnwidth]{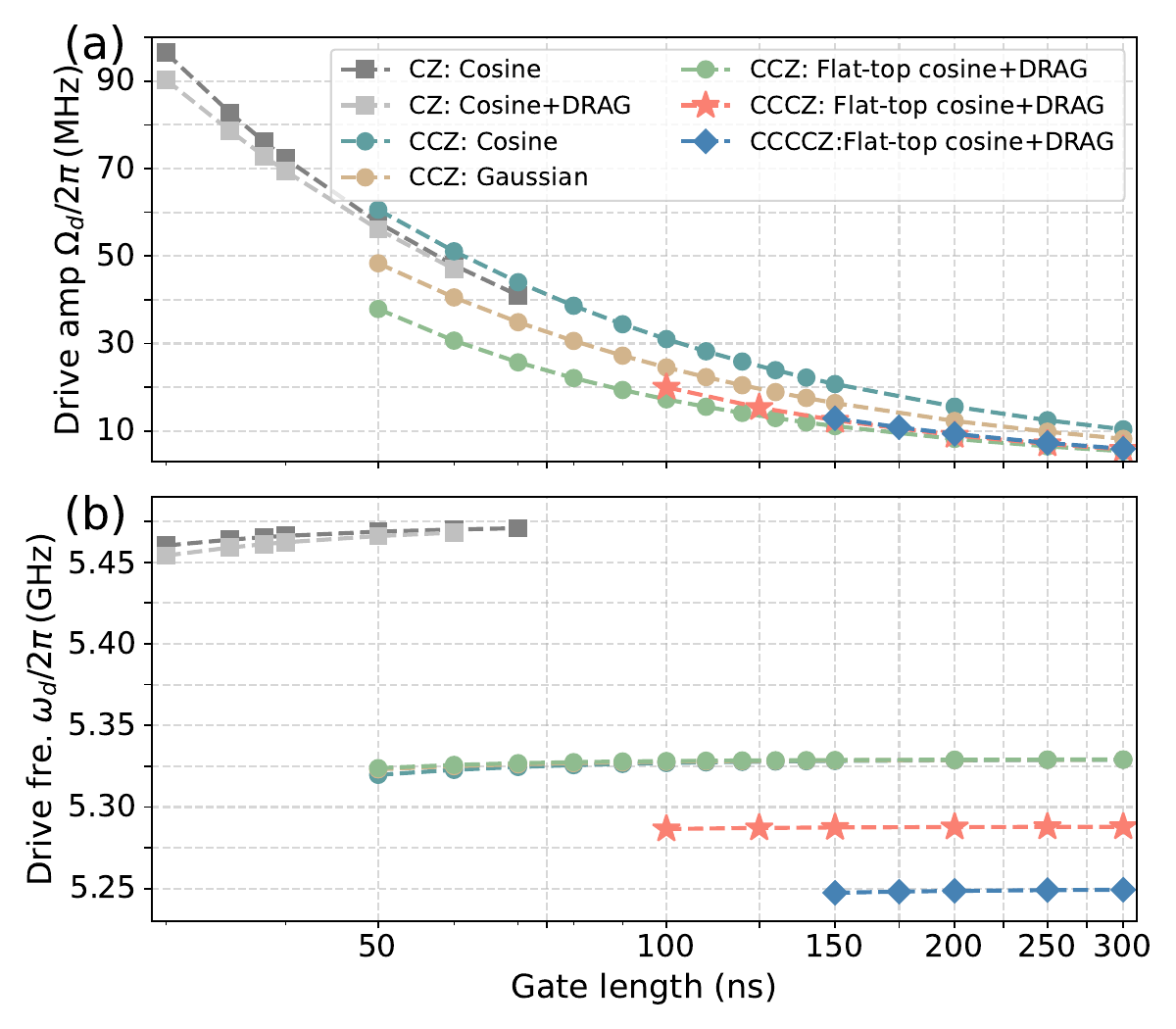}
\end{center}
\caption{The optimized $(C^{\otimes N})Z$ gate parameters (i.e., gate drive amplitudes and drive frequencies) used
for the results shown in Figs.~\ref{fig4} of the main text.}
\label{figS6}
\end{figure}

To evaluate the intrinsic gate performance (excluding decoherence effects), we employ the metric
of state-average gate fidelity~\cite{Pedersen2007}. For $(C^{\otimes N})Z$ gates, the fidelity is
calculated following, up to single-qubit Z rotations:
\begin{equation}
\begin{aligned}\label{eqC5}
F_{N}=\frac{{\rm Tr}(\tilde{U}^{\dagger}\tilde{U})+|{\rm Tr}(U_{\rm i}^{\dag}\tilde{U})|^{2}}{n(n+1)},
\end{aligned}
\end{equation}
where $\tilde{U}$ represents the truncated actual evolution operator within the $n=2^{N+1}$-dimensional
computational subspace and $U_{\rm i}$ corresponds to the ideal target gate. Using this framework, we
characterize the $(C^{\otimes N})Z$ gate performance as a function of gate length (see Fig.~\ref{fig4} of the
main text), with corresponding optimized parameters shown in Fig.~\ref{figS6}.

Note that the actual evolution operator $\tilde{U}$ is obtained through numerical simulation of the gate dynamics
using the QuTiP package~\cite{Johansson2012}, where each fluxonium is truncated to its four lowest energy levels and each transmon coupler
is modeled as an anharmonic oscillator (Eq.~\ref{eqA4}) truncated to three levels. For the CCCCZ gate simulations, we
additionally project the full five-fluxonium system Hamiltonian (with a dimension of 82,944) onto a reduced subspace containing
states whose energies lie below 24 GHz, yielding a manageable dimension of 6,096 while preserving the essential physics
of the fluxonium system.

\section{Gate errors due to relaxation and dephasing of non-computational gate levels}\label{D}

In high-coherence fluxonium systems, as demonstrated in Ref.~\cite{Ding2023}, the dominated incoherence gate error
should be from relaxation and dephasing of non-computational gate states. To evaluate these incoherence gate
errors, here we thus focus on the gate transition $|1\rangle\otimes|1,...,1\rangle\leftrightarrow|2\rangle\otimes|1,...,1\rangle$.
Following the procedure given in Refs.~\cite{Abad2023,Zhao2025}, the incoherence gate error from the relaxation
and dephasing (white noise) are
\begin{equation}
\begin{aligned}\label{eqD1}
&\epsilon_{1}=1-F_{1}=\frac{3}{32}\frac{t_{g}}{T_{1}^{21}}+\frac{13}{80}\frac{t_{g}}{T_{\phi,{\rm white}}^{21}},
\\&\epsilon_{2}=1-F_{2}=\frac{35}{576}\frac{t_{g}}{T_{1}^{21}}+\frac{11}{96}\frac{t_{g}}{T_{\phi,{\rm white}}^{21}},
\\&\epsilon_{3}=1-F_{3}=\frac{67}{2176}\frac{t_{g}}{T_{1}^{21}}+\frac{65}{1088}\frac{t_{g}}{T_{\phi,{\rm white}}^{21}},
\\&\epsilon_{4}=1-F_{4}=\frac{131}{8448}\frac{t_{g}}{T_{1}^{21}}+\frac{43}{1408}\frac{t_{g}}{T_{\phi,{\rm white}}^{21}}
\end{aligned}
\end{equation}
for $CZ$, $CCZ$, $CCCZ$, and $CCCCZ$, respectively, where $T_{1}^{21}$ and $T_{\phi,{\rm white}}^{21}$ denote the relaxation time and the dephasing (white noise) time of the non-computational level $|2\rangle\otimes|1,...,1\rangle$.

\end{document}